\documentclass[9pt,twocolumn,twoside]{pnas-new}

\usepackage[utf8]{inputenc}
\usepackage[T1]{fontenc}
\usepackage{mathptmx}
\usepackage{physics}
\usepackage{siunitx}
\usepackage[version=4]{mhchem}

\usepackage{hyperref}
\hypersetup{
    colorlinks=true,
    linkcolor=blue,
    filecolor=blue,      
    urlcolor=blue,
    citecolor=blue,
}

\usepackage{color,soul}
\usepackage{xspace}
\usepackage{xcolor}
\usepackage{tablefootnote}

\templatetype{pnasresearcharticle} 

\title{Formation of Dissipative Structures in Microscopic Models of Mixtures with Species Interconversion}

\author[a]{Thomas J. Longo} 
\author[b]{Nikolay A. Shumovskyi}
\author[c,d]{Bet\"ul Uralcan}
\author[b,e]{Sergey V. Buldyrev}
\author[a,f]{Mikhail A. Anisimov}
\author[d,1]{Pablo G. Debenedetti}

\affil[a]{Institute for Physical Sciences and Technology, University of Maryland at College Park, MD 20742, USA}
\affil[b]{Department of Physics, Boston University, Boston, MA 02215, USA}
\affil[c]{Department of Chemical Engineering and Polymer Research Center, Bogazici University, Bebek 34342, Istanbul, Turkey }
\affil[d]{Department of Chemical and Biological Engineering, Princeton University, NJ 08544, USA}
\affil[e]{Department of Physics, Yeshiva University, New York, NY 10033, USA}
\affil[f]{Department of Chemical and Biochemical Engineering, University of Maryland at College Park, MD 20742, USA}

\leadauthor{Longo} 

\significancestatement{We investigate the formation of nonequilibrium mesoscopic structures induced by the interconversion of species in three physically distinct microscopic binary-mixture models. We show that these models exhibit identical behavior depending on the interplay between diffusion, natural interconversion, and forced interconversion. This behavior is well-captured by a non-equilibrium thermodynamic theory. The phenomenon (referred to as ``microphase separation'') is one of the simplest examples of steady-state dissipative structures in condensed matter, such as hydrodynamic instabilities and bifurcations in chemical reactions. Nonequilibrium microphase separation may exist in chemical and biological systems, where the alternative molecular or supramolecular states may interconvert via mechanisms like polymerization, protein folding-unfolding, and self-assembly, in which the nonequilibrium conditions are imposed by an external flux of matter or energy.}

\authordeclaration{The authors declare no competing interests.}
\correspondingauthor{\textsuperscript{1}To whom correspondence should be addressed. E-mail: pdebene@princeton.edu}

\keywords{Molecular Interconversion $|$ Phase Separation $|$  Liquid Polyamorphism $|$ Dissipative Structures} 

\begin{abstract}
The separation of substances into different phases is ubiquitous in nature and important scientifically and technologically. This phenomenon may become drastically different if the species involved, whether molecules or supramolecular assemblies, interconvert. In the presence of an external force large enough to overcome energetic differences between the interconvertible species (forced interconversion), the two alternative species will be present in equal amounts, and the striking phenomenon of steady-state, restricted phase separation into mesoscales is observed. Such microphase separation is one of the simplest examples of dissipative structures in condensed matter. In this work, we investigate the formation of such mesoscale steady-state structures through Monte Carlo and Molecular Dynamics simulations of three physically distinct microscopic models of binary mixtures that exhibit both equilibrium (natural) interconversion and a nonequilibrium source of forced interconversion. We show that this source can be introduced through an internal imbalance of intermolecular forces or an external flux of energy that promotes molecular interconversion, possible manifestations of which could include the internal nonequilibrium environment of living cells or a flux of photons. The main trends and observations from the simulations are well captured by a non-equilibrium thermodynamic theory of phase transitions affected by interconversion. We show how a nonequilibrium bicontinuous microemulsion or a spatially modulated state may be generated depending on the interplay between diffusion, natural interconversion, and forced interconversion.
\end{abstract}

\dates{This manuscript was compiled on \today}
\doi{\url{www.pnas.org/cgi/doi/10.1073/pnas.XXXXXXXXXX}}

\begin{document}

\maketitle
\thispagestyle{firststyle}
\ifthenelse{\boolean{shortarticle}}{\ifthenelse{\boolean{singlecolumn}}{\abscontentformatted}{\abscontent}}{}


\dropcap{T}wo distinct molecular species may separate if the interactions between them are energetically unfavorable relative to the interactions between like species. The most recognizable example is the almost complete separation of water and oil. During phase separation via spinodal decomposition (mixture quenched into the unstable region) transient patterns of mesoscale inhomogenieties are observed \cite{Vvedensky_Transformations_2019,Fultz_Phase_2020}. However, such patterns are unstable and disappear in equilibrium, although they may be ``frozen'' by rapid quenching, as commonly observed in glasses \cite{Kim_Metallic_2013,Lin_Nanocrystallization_2015}. Alternatively, in equilibrium, examples of mesoscale structures are present in bicontinuous or spatially modulated microemulsions \cite{Gompper_Lattice_1994,Andelman_Modulated_2009} and microphase separation of diblock or polyelectrolyte copolymers \cite{Jones_Soft_Matter_2002,Borukhov_Polyelectrolyte_2000}, where these mesoscale patterns are the result of the minimization of the equilibrium free energy \cite{anisimov_thermodynamics_2018,li_non-equilibrium_2020}.

In this work, we consider three physically distinct microscopic realizations of a binary mixture, where the alternative species naturally interconvert according to thermodynamic equilibrium. We show that the presence of a non-equilibrium force (either originating internally or imposed externally), which, if large enough, can cause the alternative species to be present in equal amounts (``forced interconversion''), drives the system away from equilibrium and produces the striking phenomenon of steady-state, restricted (``microphase'') separation into mesoscale domains. We observe that the structure of the phase domains formed from nonequilibrium microphase separation resembles modulated or bicontinuous microemulsion structures \cite{Gompper_Lattice_1994,Andelman_Modulated_2009,Jones_Soft_Matter_2002,Borukhov_Polyelectrolyte_2000,anisimov_thermodynamics_2018,li_non-equilibrium_2020}. However, contrary to the patterns formed in equilibrium or in metastable (``frozen'') conditions, these nonequilibrium structures persist in steady-state due to the continuous energy supply. Thus, steady-state microphase separation is one of the simplest examples of dissipative structures in condensed matter. The characteristic length scale of this dissipative structure emerges as a result of the competition between forced interconversion and phase growth. If the source of forced interconversion is not sufficiently strong to overcome the natural interconversion of alternative species, then the phenomenon of phase amplification, the growth of one stable phase at the expense of another phase, is observed \cite{Ricci_Computational_2013,Shum_Phase_2021,MFT_PT_2021}. In fluids that exhibit molecular interconversion of species, the conservation of the number of alternative molecules is broken and phase amplification would occur to avoid the formation of an energetically unfavorable interface. However, in a nonequilibrium system in the presence of an external source of energy, the formation of mesoscale interfaces (microphase separation) may become favorable \cite{MFT_PT_2021}.

An external-forced interconversion source can be achieved in physical systems through the interactions of energy-carrying particles, such as photons, that may break intramolecular bonds \cite{Miyata_PolyemrMix_2017}. A possible source could also be the nonequilibrium environment of biological cells, where the associated continuous dissipation of energy can be used to drive chemical reactions \cite{Weber_Active_2019,Boeynaems_Protein_2018,Hyman_LLBiology_2014}, or it could be achieved chemically through an external flux of matter or heat \cite{huberman_striations_1976,Verdasca_Chemically_1995,carati_chemical_1997}. In simulations, the nonequilibrium state could be achieved through an internal imbalance of intermolecular forces or an imbalance of the free energy by introducing an external source of energy. Previous studies of a nonequilibrium phase-separating binary mixture in the presence of an external reaction source, originally introduced by Glotzer \textit{et al}.  \cite{glotzer_monte_1994,Glotzer_consistent_1994,glotzer_reaction-controlled_1995,lefever_comment_1995,Glotzer_Response_1995,Glotzer_Review_1995,Christensen_Segregation_1996,Krishnan_Molecular_2015,Lamorgese_Spinodal_2016}, as well as a more recent dissipative chiral model of interconverting enantiomers with unbalanced intermolecular forces \cite{Uralcan_Interconversion_2020} and a nonequilibrium hybrid Ising/lattice-gas model with an imbalance of internal energy \cite{Longo_Structure_2021}, have all demonstrated steady-state microphase separation.


Previous theoretical studies of phase separation in chemically-reactive binary mixtures have demonstrated that the formation of steady-state dissipative structures (in systems with a simple $\ce{A <=> B}$ interconversion reaction) is only possible under nonequilibrium conditions \cite{carati_chemical_1997,Lamorgese_Spinodal_2016,li_non-equilibrium_2020}. In their seminal work, utilizing a scalar field theoretical approach, Li and Cates have shown that different kinds of steady-state structures may be formed in a nonequilibrium system with a combination of diffusion and chemical-reaction dynamics  \cite{li_non-equilibrium_2020,Heirarchical_Li_2021}. In our work, we study the formation of nonequilibrium dissipative structures by investigating three nonequilibrium microscopic models of phase-separating binary systems with mixed dynamics. We unify the behavior of the three distinct models through a theoretical approach, conceptually similar to the study of Li and Cates. In our case, however, we explicitly consider the evolution of the three models towards equilibrium and their behavior at equilibrium. We describe the interconverting mixtures phenomenologically by a single order parameter that possesses both conserved (diffusion) and nonconserved (interconversion) dynamics (see Sec. 1.C in the SI for more details). This dynamic feature of the order parameter is inherent to the three atomistic models considered in our work. In our approach, we separate the effect of interconversion into two parts: an equilibrium (natural) interconversion, governed by the same free energy as phase separation, and a nonequilibrium (forced) interconversion, controlled by an external-energy source.


The microscopic models are simulated through Monte Carlo (MC) and Molecular Dynamics (MD) methods. We consider a hybrid-lattice (HL) model with an externally introduced imbalance of internal energy \cite{Longo_Structure_2021}, a dissipative chiral mixture model (DCM) with unbalanced intramolecular forces \cite{Latinwo_MolecModel_2016,Uralcan_Interconversion_2020}, and a hard-core-shoulder (HCS) model with an external energy source. We also qualitatively compare the behavior of the simulated models with the behavior of Glotzer \textit{et al.}'s nonequilibrium binary-mixture model \cite{glotzer_monte_1994,Glotzer_consistent_1994,glotzer_reaction-controlled_1995,Glotzer_Review_1995,Christensen_Segregation_1996,Krishnan_Molecular_2015,Lamorgese_Spinodal_2016}. We show that all of these models can be described through the same theoretical approach \cite{MFT_PT_2021} and we obtain {a} quantitative agreement between the theory and simulations. We show that during the evolution of the system to a steady state, the interplay between phase separation, natural interconversion, and forced interconversion may generate novel dissipative structures – a nonequilibrium bicontinuous microemulsion, revealed by MD, or a nonequilibrium spatially modulated state, as observed in MC studies. We show that under certain constrains the three models with different origins of interconversion [\textit{i.e.} our three microscopic models: HL \cite{Shum_Phase_2021,Longo_Structure_2021}, DCM \cite{Latinwo_MolecModel_2016,Uralcan_Interconversion_2020}, and HCS, as well as the model of Glotzer \textit{et al.}  \cite{glotzer_monte_1994,Glotzer_consistent_1994,glotzer_reaction-controlled_1995,Glotzer_Review_1995,Christensen_Segregation_1996,Krishnan_Molecular_2015,Lamorgese_Spinodal_2016}] may exhibit identical behavior. 

\section{Simulated Models}
We simulate three microscopic models: a nonequilibrium hybrid-lattice (HL) model with an externally induced imbalance of internal energy, a dissipative chiral-mixture (DCM) model where the internal source is {an imbalance} of intermolecular forces, and a hard-core-shoulder (HCS) model where the external source is caused by energy carrying agents, such as photons, neutrons, or adenosine triphosphate (ATP) molecules. In each model, the alternative species interconvert naturally (according to thermodynamic equilibrium), but also, forcefully via either an internal or external source of energy or matter.

\subsection{Hybrid-Lattice (HL) Model with an Imbalance of Internal Energy} This model was introduced, in its equilibrium formulation, in ref. \cite{Shum_Phase_2021}, and the nonequilibrium formulation was introduced, and qualitatively considered, in ref. \cite{Longo_Structure_2021}. To model diffusion and interconversion dynamics in a binary system, we consider a simple lattice model where the particles of different types are represented by spins of different orientations. Diffusion is simulated by ``swapping'' a pair of randomly selected neighboring spins and interconversion is simulated by ``flipping'' a spin at a randomly selected lattice cite. At each MC step the probability that a random spin will attempt to flip is $p_r$, while swapping a randomly selected pair of nearest neighbor spins will be attempted with probability $1-p_r$. This step is accepted according to the standard Metropolis criterion \cite{Metropolis_Basic_1963}.

In the nonequilibrium formulation of the HL model, an additional energy, $E$, due to the external source of {forced} interconversion is incorporated into the Boltzmann factor for the Metropolis criterion of a spin flip as $p \sim \exp{[-(\Delta U -E)/k_\text{B}T]}$, where $\Delta U$ is the difference in internal energy of the system for this step. When $E=0$, the system evolves according to the equilibrium formulation detailed in ref. \cite{Shum_Phase_2021}, which leads to phase amplification. If $E>0$ and is large enough to overcome natural interconversion (and, consequently, phase amplification), then steady-state microphase separation occurs. The energy source, $E$, reduces the thermodynamic energy barrier between inhomogeneous and phase separated states. Thus, it promotes interconversion to an equal composition of species in opposition to natural interconversion where, in general, the relative population of interconverting species varies with thermodynamic conditions. For energy $E \ge 12$, the external source of forced interconversion is always greater than $\Delta U$ (cubic lattice in 3-d), such that spin interconversion occurs with probability, $p_r$, and the Metropolis criterion is always accepted ($p\sim 1$). This scenario makes this model equivalent to the model of Glotzer \textit{et al.} \cite{glotzer_monte_1994,Glotzer_consistent_1994,glotzer_reaction-controlled_1995,Glotzer_Review_1995}. For $E<12$, the local environment of the selected species influences the probability of the interconversion reaction occurring ($p<1$). Thus, whether a spin-flip will be accepted at each MC step is determined by both the spin-flip probability, $p_r$, and the Metropolis criterion, $p$. For the diffusion dynamics, the spin-swap step is performed according to the standard Metropolis criterion without any additional energy source, such that diffusion is the same in both the equilibrium and nonequilibrium formulations of this model. For more information about the HL model see ref. \cite{Longo_Structure_2021}.

\subsection{Dissipative Chiral Mixture (DCM) Model with Unbalanced Forces}
The chiral-tetramer force-field model is based on a 4-site flexible molecule that can adopt two mirror-image configurations \cite{Latinwo_MolecModel_2016,Petsev_Effect_2021,Uralcan_Interconversion_2020} and is simulated via MD. In the conservative formulation, all intermolecular forces are balanced by taking into account the multi-body forces arising from an explicit chirality-dependent characteristic interaction energy term, as detailed in ref.~\cite{Petsev_Effect_2021}. In this case, phase amplification is observed where the growth of one of the two alternative states is restricted only by system size. The dissipative formulation, on the other hand, exhibits an imbalance of intermolecular forces resulting from not applying the gradient operator to the chirality dependent term in the potential energy function \cite{Uralcan_Interconversion_2020} (see Sec. 2.B in the SI for details). The phenomenon of microphase separation is observed in the dissipative formulation of this model. Now, in this work, we show that the ubiquitous nature of this striking phenomenon can be better understood in connection with other nonequilibrium models with molecular interconversion.

In the DCM model, molecules feature left- and right-handed configurations. Interconversion from one enantiomer to another is controlled through bond rotations about the dihedral angle (see Fig. S3 in the SI for details). The force constant for dihedral rotation controls the kinetics of enantiomeric interconversion, and is denoted by $k_d$. The model includes a chiral bias parameter, $\lambda$, whose sign defines the nature of chiral bias, such that: $\lambda < 0$ favors local (short-ranged) heterochiral interactions, $\lambda > 0$ favors homochiral interactions (enantiopure states), and $\lambda = 0$ represents a bias-free scenario \cite{Latinwo_MolecModel_2016}. For the simulations considered in this work, $\lambda=0.5$ (energetically favored local interactions between molecules of the same chirality). 

\begin{figure*}[t]
    \centering
    \includegraphics[width=0.32\linewidth]{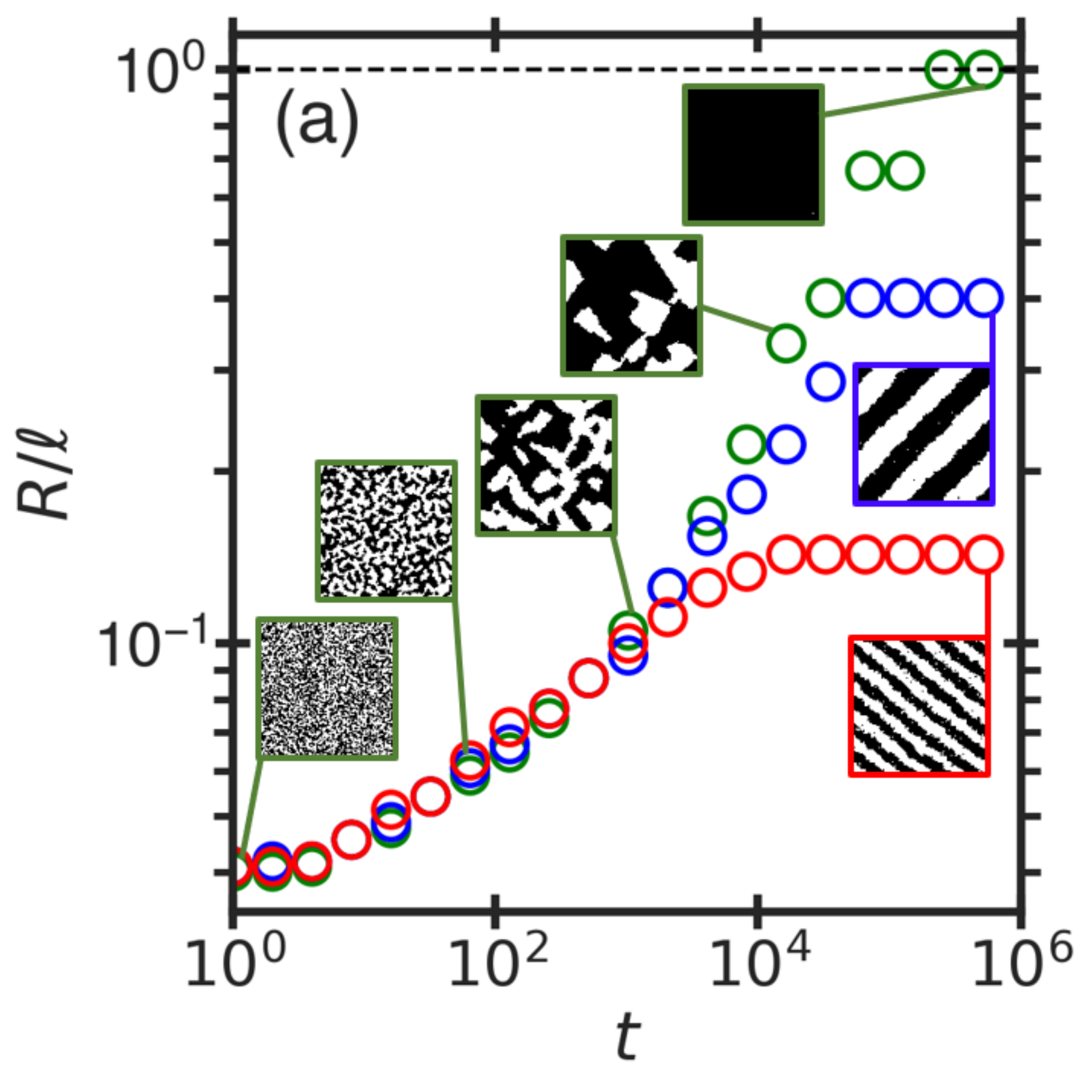}
    \includegraphics[width=0.32\linewidth]{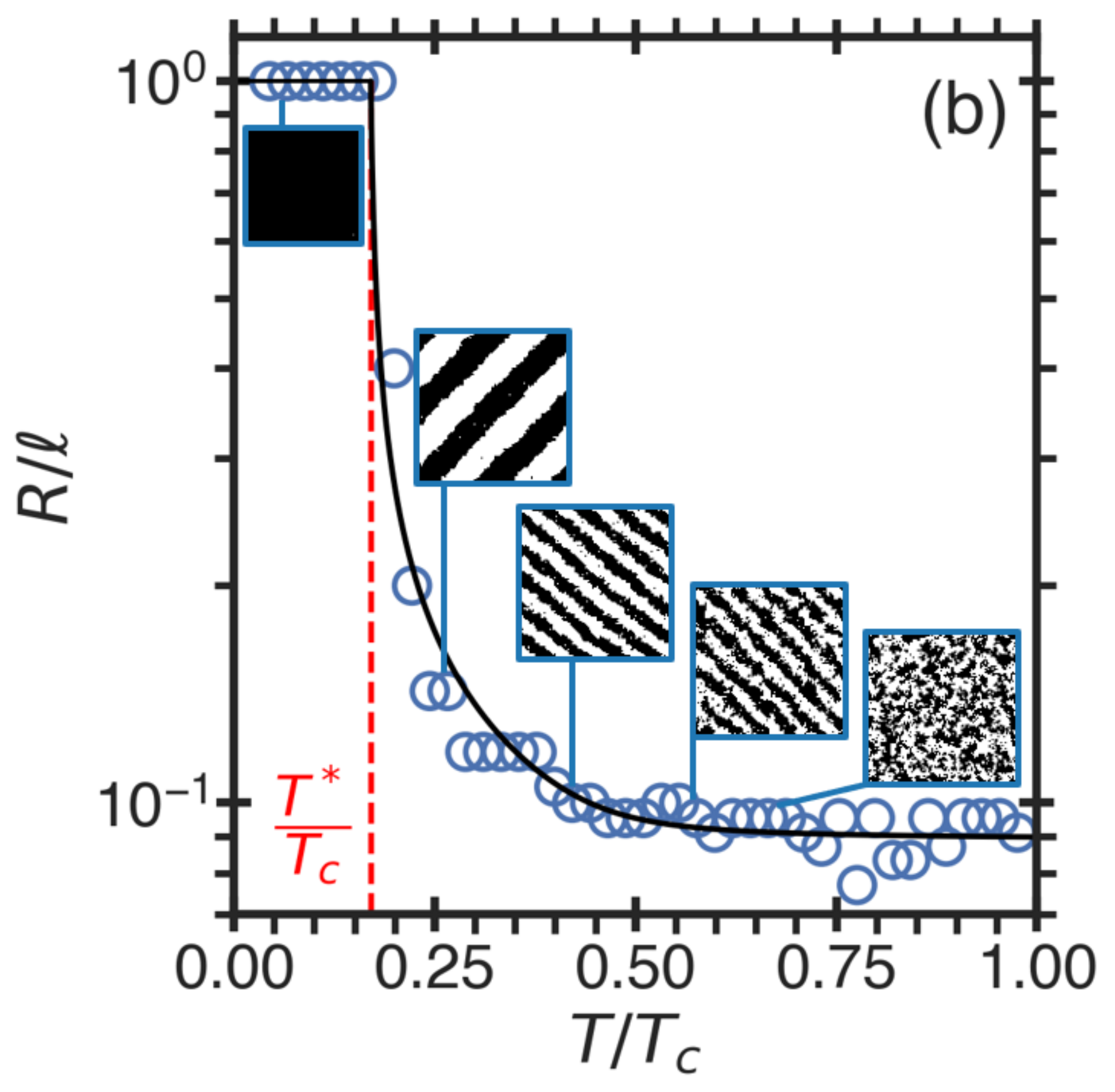}
    \includegraphics[width=0.32\linewidth]{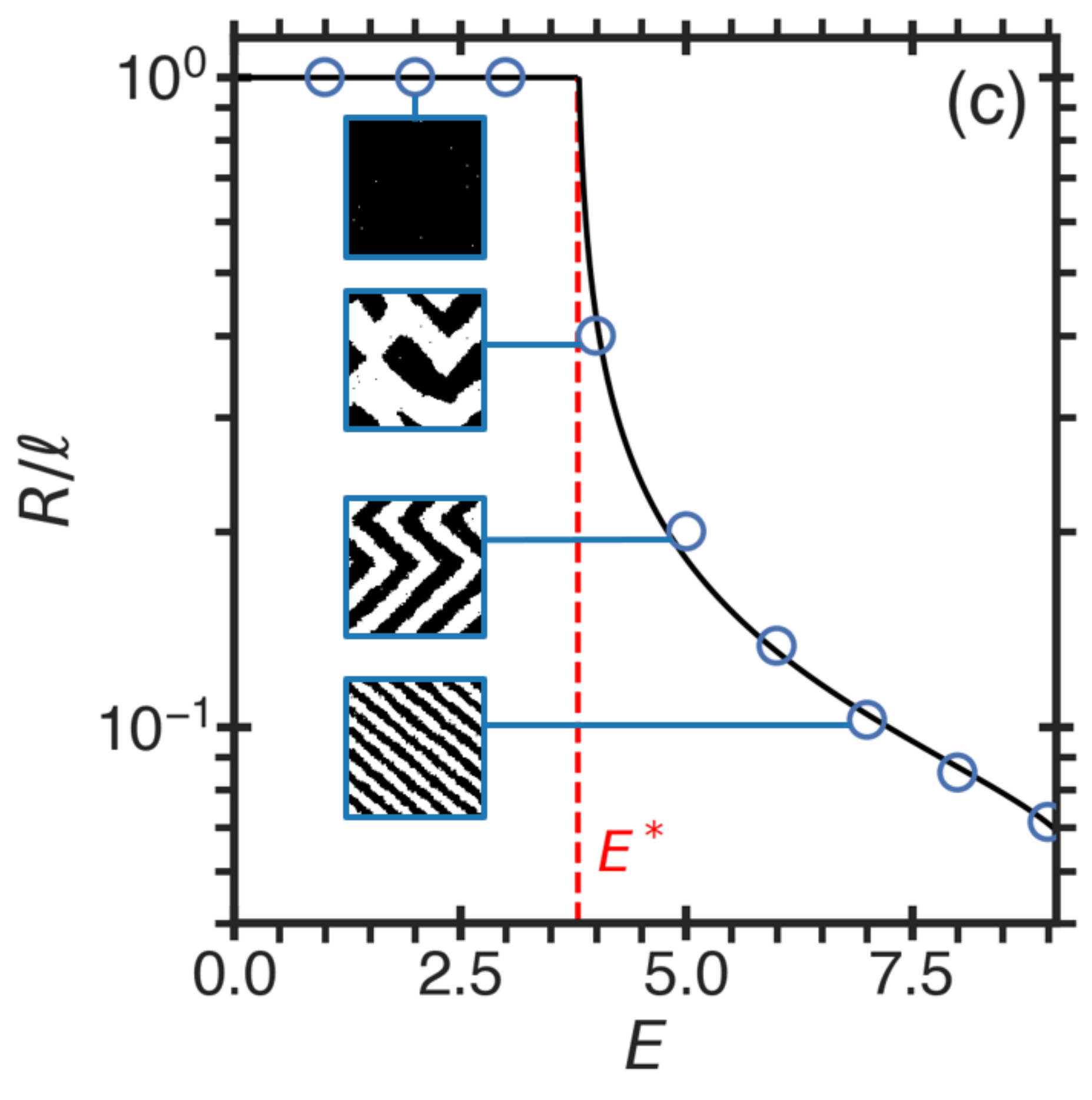}
    \caption{Effect of forced interconversion on domain size, $R$, normalized by the system size, $\ell$, in the HL model. (a) The time dependence of the domain growth for energy source $E=5$ and interconversion probability $p_r = 1/128$ at $T/T_\text{c} = 0.24$ (green), $T/T_\text{c} = 0.27$ (blue), and $T/T_\text{c} = 0.40$ (red), where $T_\text{c}= 4.511$ \cite{Heuer_Critical_1993}. The horizontal dashed line indicates the size of the simulation box, $R=\ell$. (b) {Temperature} dependence of the steady-state domain size for $E=5$ and $p_r = 1/128$. The vertical dashed line indicates the onset temperature, $T^*/T_\text{c}$. (c) Dependence of the steady-state domain size on the energy of {forced} interconversion for $p_r = 1/256$ and $T/T_\text{c} = 0.31$. The vertical dashed line denotes the onset source energy, $E^*$. In (a-c), the system is simulated on a 3-dimensional lattice of size $\ell = 100$. The open circles are the results of MC simulations, the images are snapshots of the system for selected conditions, and the curves are the theoretical predictions (see Sec. 2.A in the SI for details). In (a-c), black denotes up-spins and white denotes down-spins. }
    \label{Fig_Theory_NoneqHybrid}
\end{figure*}

\subsection{Hard-Core-Shoulder (HCS) Model with an External Energy Source}
Unlike the previous examples, this model, which has not been considered before, utilizes a tunable source of forced interconversion, implemented through the interactions with an external flux of energy carrying agents. In biological systems, these agents can be thought of as ATP molecules, which change the conformation of a protein between two phase-segregating states. We consider a system initially consisting of an equal number of $N_A=N_B=N/2=32,000$ identical hard spheres of type A and B with diameter $\sigma$. While all particles repel each other as hard spheres with diameter $\sigma$, phase segregation is generated by the additional repulsion between A and B particles via a square shoulder potential at distance $d=1.3\sigma$, with energy $\epsilon_0$. We simulate the interactions within the system via event-driven MD with discontinuous potentials \cite{Alder_Studies_1959,Rapaport_Art_2004}. Additionally, we introduce $N_\text{ag}=10,000$ agents as an external source of energy, which collide with particles A and B at a distance $b<\sigma$. The agents contain an additional energy $\epsilon$ (measured in units of $\epsilon_0$), which upon collision with particles A or B boosts the probability of species interconversion (\ce{A <=>B}). Physically, $\epsilon$ can be regarded as an external energy carried by an ATP molecule or another active agent, which can compensate the effect of the heat of mixing arising from interconversion. In our simulations, this reaction occurs instantaneously, without any metastable intermediate state of either species.

The systems considered in this work are simulated in an $\ell\times\ell\times\ell$ box of length $\ell=40\sigma$ with periodic boundaries at temperature $T$, measured in units of $\epsilon_0/k_\text{B}$. To regulate the temperature, we use a Berendsen thermostat \cite{Berendsen_Molecular_1984}. The collision of the agents with species A or B occurs with conservation of linear and angular momentum of the pair as well as with total potential energy change, $\Delta U$. The total energy is composed of potential, kinetic, and external energy, in which the external energy is incorporated into the kinetic energy of the colliding particles. The equilibrium formulation occurs with conservation of energy, such that $\epsilon=0$. 

In the equilibrium formulation, the agents either contain no additional energy, $\epsilon$, or the cross-sectional area of their interaction with the species, $b^2$, is zero, such that the agents pass through species A or B without interacting. Physically, this corresponds to a scenario when the energy is unable to transfer from the agents to the species in the system. In the nonequilibrium formulation, the agents possess both the additional energy and cross-sectional area necessary to interact and convert species A to B and vice-versa. Without an external source of energy, an energetically costly interconversion reaction violates the conservation laws, so that the particles will recoil and interconversion will not happen. However, in the presence of an external source of energy, provided by the agents, the interconversion reaction may happen both in favorable or unfavorable conditions, just as in the HL model.  

The particles (A, B, and the agents) have equal masses $m$, and the simulation time is measured in units of  $\sigma\sqrt{m/\epsilon_0}$. Overall, in the HCS model, the system is a dense fluid of hard spheres with a molecular mobility, $M\propto \sqrt{T}$ (see Fig.~S4), and interconversion rate proportional to the interaction cross section, $b^2$ (see Sec. 2.C in the SI for details).

\section{Simulation Results}
In this section, we discuss the conditions for the formation of dissipative structures using as parameters{:} the distance to the equilibrium critical temperature of demixing, $T_\text{c}$, the interconversion probability ($p_r$, $k_d^{-2}$, or $b^{2}$ respectively), and source strength ($E$ or $\epsilon$ respectively). We show the ubiquitous nature of the formation of dissipative structures in the different simulated models.

\subsection{HL Model}
The effect of the source of forced interconversion is introduced into the HL model through a tunable imbalance in the internal energy via the energy of forced interconversion ($E$), such that the source boosts the probability for two alternative species to interconvert into equal amounts. If forced interconversion is not strong enough to overcome natural interconversion (which corresponds to equilibrium conditions), then phase amplification (the growth of one stable phase at the expense of the other) is observed, provided that the natural interconversion probability is significant. If forced interconversion overcomes natural interconversion by a sufficient amount, we observe that the locally phase-separated domains stop growing upon reaching a characteristic size, as illustrated in Fig.~\ref{Fig_Theory_NoneqHybrid}a. We define the temperature and energy at which this occurs as the ``onset'' of microphase separation ($T^*$ and $E^*$). As shown in Figs.~\ref{Fig_Theory_NoneqHybrid}b and \ref{Fig_Theory_NoneqHybrid}c, for temperatures and energies, $T>T^*$ and $E>E^*$ (for a given probability, $p_r$), dissipative structures are observed and the steady-state domain size decreases as the energy of {forced} interconversion increases and as the temperature becomes closer to the critical demixing temperature, $T_\text{c}$. 

We find that the dissipative domain structure has the form of spatially-modulated stripes due to the symmetry and boundary conditions of the lattice on which the MC simulations are performed. As shown by the simulation snapshots in Fig.~\ref{Fig_Theory_NoneqHybrid}b, the striped pattern becomes more disordered when the domain size becomes comparable with the correlation length of concentration fluctuations (see Eq.~S6). We also observed that the chance of forming defects, like kinks or corners in the phase-domain structure, increases for larger-sized microphase structures (see the simulation snapshots in Fig.~\ref{Fig_Theory_NoneqHybrid}c). Only after an astronomically large simulation time will these kinks fully diffuse to produce the final steady-state structure. We attribute this effect to the fact that the energy penalty from forming a corner is minuscule when compared to the bulk, such that these defects would take a longer time to diffuse. Yet, for smaller microphase domains, these defects generate a larger energy penalty when compared to the bulk, and thus, are removed faster.

\subsection{DCM Model}
\begin{figure}[t]
    \centering
    \includegraphics[width=\linewidth]{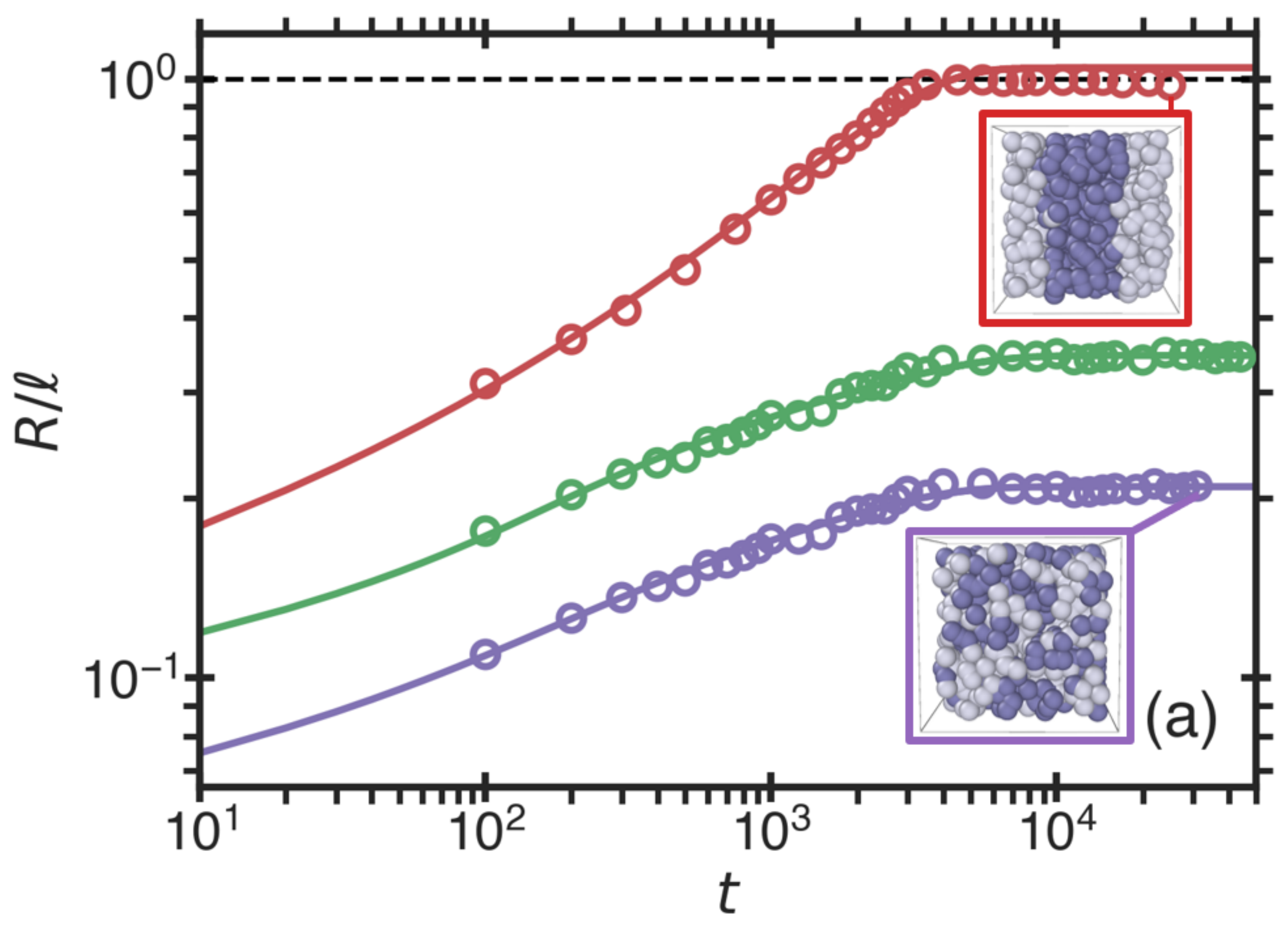}
    \includegraphics[width=\linewidth]{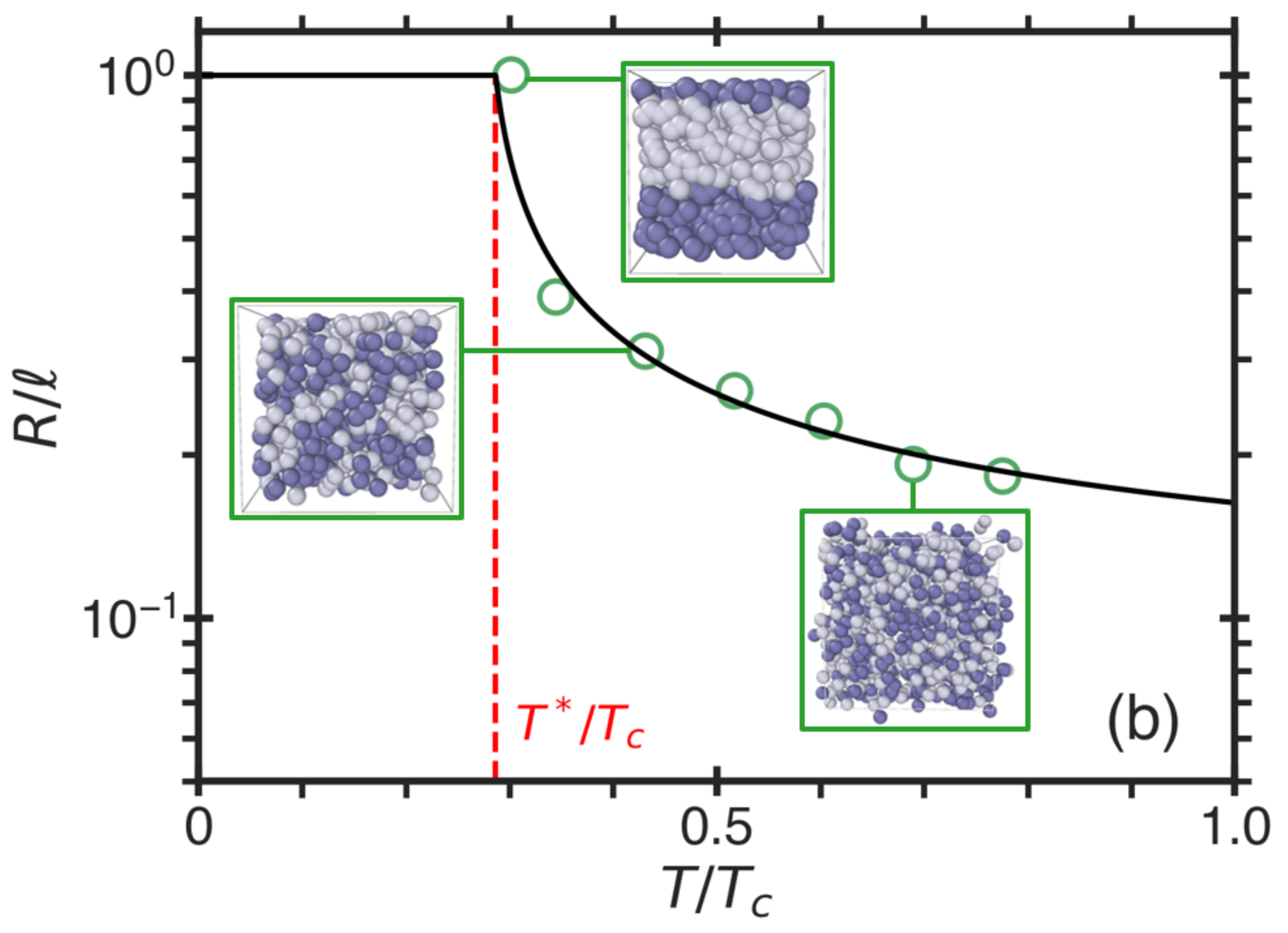}
    \caption{Steady-state domain size, $R$, normalized by the size of the system, $\ell$, in the DCM model. (a) The time evolution of the domain size for different interconversion rates, $\sim T^2/k_d^2$ \cite{Uralcan_Interconversion_2020}, tuned by the rigidity parameter, $k_d$, as $k_d=3$ (purple), $k_d=5$ (green), and $k_d=9.86$ (red) at the reduced pressure $P=0.1$ and $T/T_\text{c}=0.35$, where $T_\text{c} (P=0.1)=2.32$ \cite{Uralcan_Interconversion_2020}. (b) The {normalized} steady-state domain size as a function of temperature at $P=0.1$ and $k_d=5$. The vertical dashed line indicates the onset temperature, $T^*/T_\text{c}$. In (a) and (b), the open circles correspond to simulation results, the curves correspond to the theoretical predictions (see Sec. 2.B and Table S.3 in the SI for details), and the images show snapshots of the system simulated at the indicated conditions. In (a-b), dark/clear spheres correspond to the L-/D-configuration of a chiral tetramer (spheres are located a tetramer's center of mass).}
    \label{Fig_Chiral}
\end{figure}

In the DCM model, the nonequilibrium condition is mimicked internally through an imbalance in intermolecular forces \cite{Petsev_Effect_2021,Uralcan_Interconversion_2020}. This imbalance is introduced if the chirality-dependent characteristic energy scale is not included (as it should) when applying the gradient operator to calculate site-site forces. As illustrated in Fig.~\ref{Fig_Chiral}a, the size of the microphase domains is restricted proportionally to the dihedral-force constant ($k_d$) at fixed temperature and pressure \cite{Uralcan_Interconversion_2020}, before it reaches the size of the computational box, $\ell\sim 1/q^*(k_d^*)$. Just like the nonequilibrium HL model, the domain size at fixed dihedral-force constant decreases with increasing temperature as shown for $k_d = 5$ in Fig.~\ref{Fig_Chiral}b. As depicted by the simulation snapshots in Fig.~\ref{Fig_Chiral}, the onset temperature $T^*$ and $k_d^*$, correspond to the conditions where the domain size reaches the size of the simulation box.

Unlike the HL model, where the source of forced interconversion ($E$) is uncoupled from the interconversion probability ($p_r$), in the DCM model, both of these effects are controlled by the rigidity parameter, $k_d$. Consequently, when $k_d\to\infty$ (corresponding to $p_r\to 0$ in the HL model) no interconversion, either natural or forced, occurs and the system would phase separate via Cahn-Hilliard diffusion-induced spinodal decomposition \cite{MFT_PT_2021}. A study of the equilibrium HL model demonstrated that the system, for low interconversion probability, commonly enters a metastable state with an interface rather than undergo phase amplification \cite{Shum_Phase_2021}. This phenomenon is also relevant to the DCM model, where it is increasingly likely for high enough $k_d$ that the system will enter a long-living transient state with two phases separated by an interface. This state would eventually converge to a steady-state configuration with the lowest possible interfacial energy (a single phase formed via phase amplification), for low $T$ and high $k_d$ (which corresponds to low $p_r$ in the HL model). This transient dissipative structure is depicted in the simulation snapshots of Fig.~\ref{Fig_Chiral}, where it is observed that an interface has formed between the two phases. 

\subsection{HCS Model}
\begin{figure}[t]
    \centering
    \includegraphics[width=\linewidth]{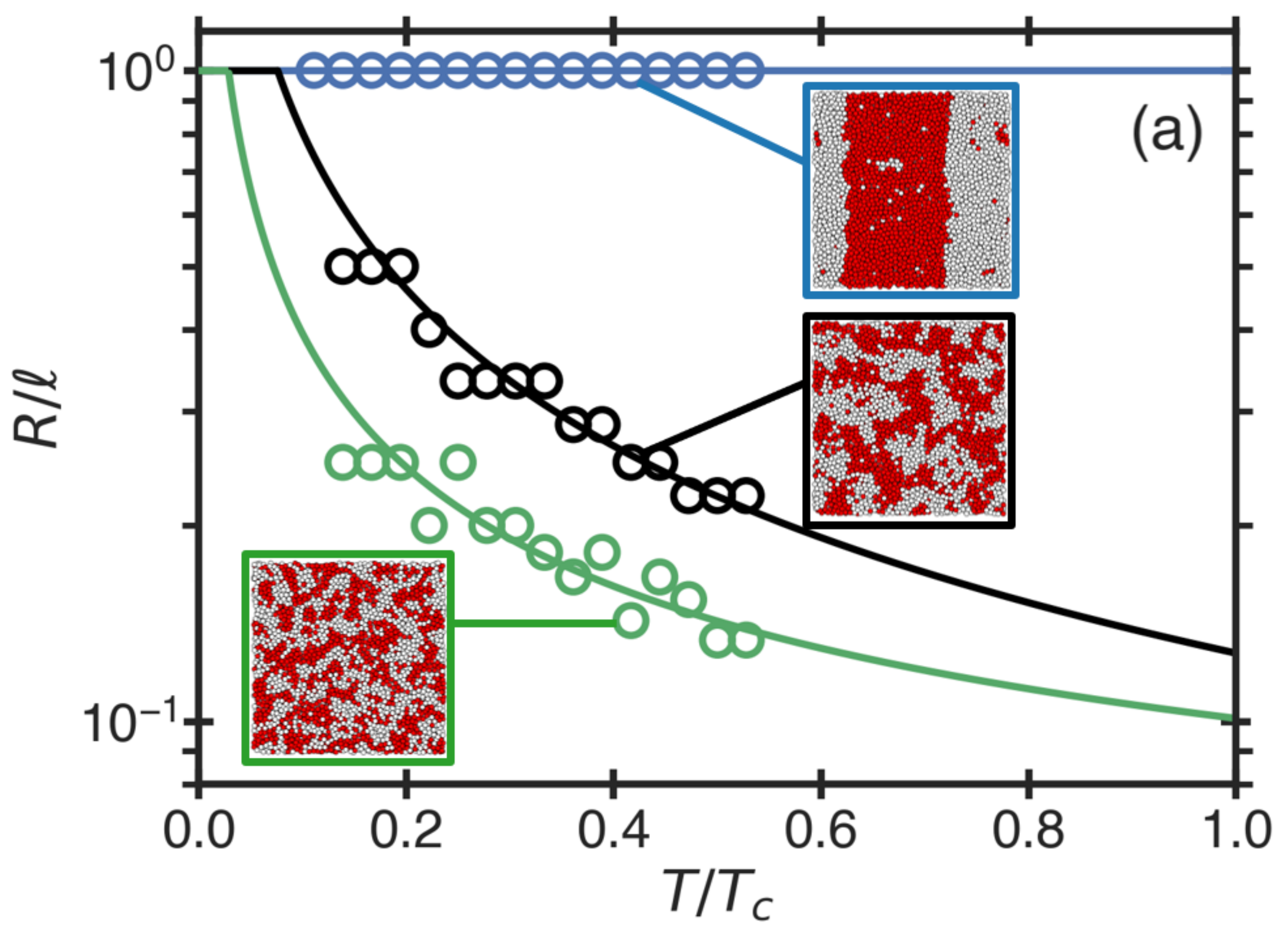}
    \includegraphics[width=\linewidth]{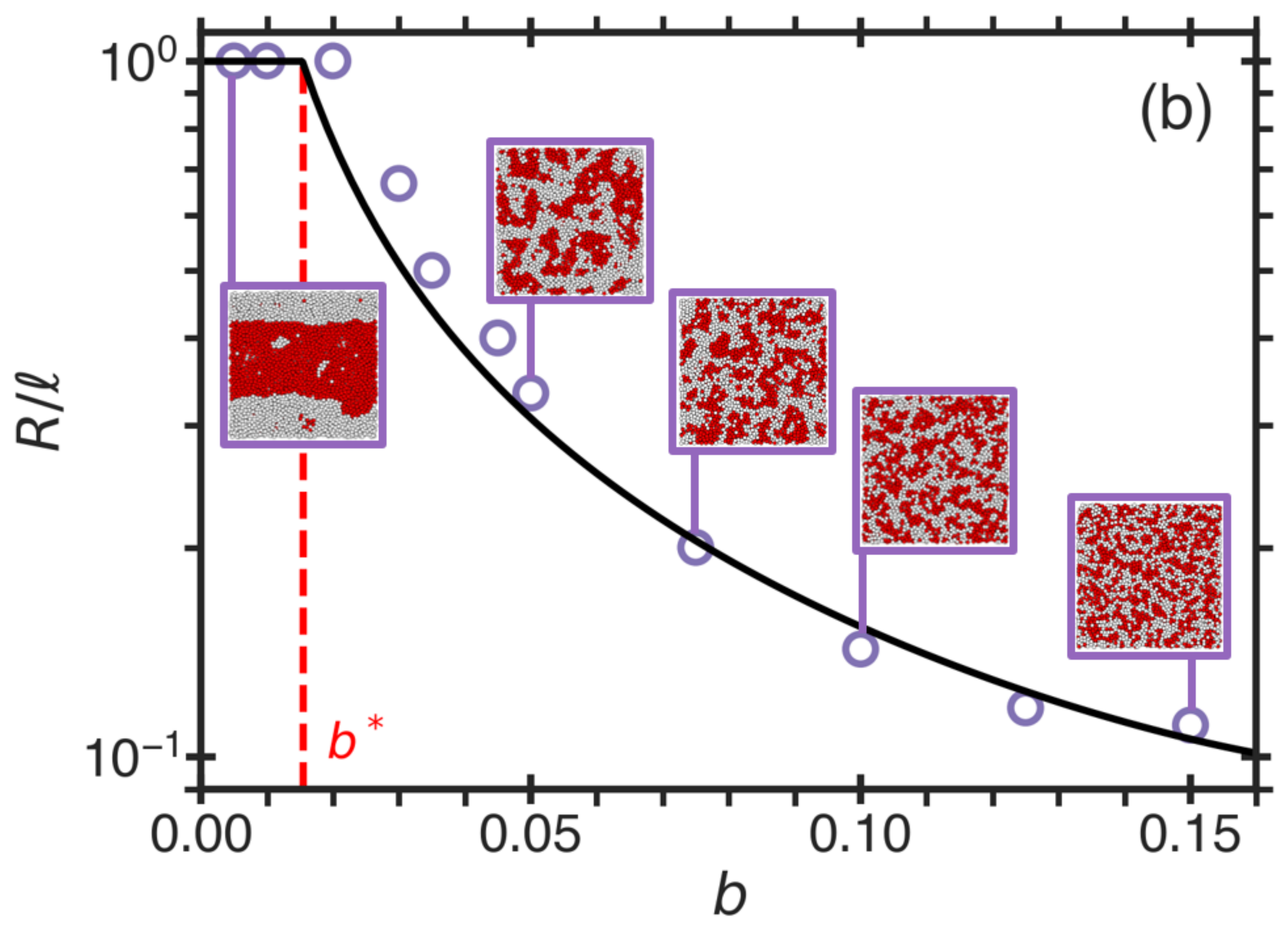}
    \caption{Steady-state domain size, $R$, normalized by the size of the system, $\ell$, in the HCS model. (a) The temperature-dependence of the {normalized} steady-state domain size for $b = 0.005$ (blue), $b = 0.050$ (black), and $b = 0.075$ (green) at $\epsilon = 10$. {(b)} The normalized steady-state domain size as a function of interconversion probability, $b$, for the energy source $\epsilon = 12$ and $T/T_\text{c}=0.22$, where $T_\text{c} = 3.6\pm 0.05$ (See Fig.~S5 for details). The vertical dashed line indicates the onset interconversion probability, $b^*$. In (a) and (b), the open circles correspond to simulation of $64,000$ particles, the curves correspond to the theoretical predictions (see Sec. 2.C. in the SI for details), and the images show snapshots of the system simulated at the indicated conditions.}
    \label{Fig_DomainSize_ShoulderModel}
\end{figure}

In the HCS model, non-relativistic energy-carrying particles are introduced as a source of forced interconversion. They carry energy, $\epsilon$, and transfer this energy via molecular collisions with cross-sectional area $b^2$. When the additional particles carry no extra energy, $\epsilon = 0$, only natural interconversion, with a probability $b$, will occur. However, due to the external source of energy provided by the particles, forced interconversion will occur just like in the previously considered models. Similar to those models, in the HCS model, the characteristic domain size decreases as a function of temperature and interconversion probability, $b$, as depicted in Fig.~\ref{Fig_DomainSize_ShoulderModel}. In this case, when $b = 0$, then regardless of $\epsilon$, no interconversion will be possible. For conditions $b < b^*\approx 0.02$ at $T < T^* = 0.22T_\text{c}$ and $\epsilon<\epsilon^* = 10$, corresponding to the onset of microphase separation, the system enters a transient state with an interface, similar to the DCM model, as illustrated by the simulation snapshots in Fig.~\ref{Fig_DomainSize_ShoulderModel}b. 

We note that below the onset of microphase separation, the characteristic steady-state domain size is comparable to the size of the simulation box, $\ell\sim 1/q^*$. Consequently, the onset conditions for all models depends on the system size. In addition, for small system sizes phase amplification occurs faster than for large systems, such that instead of entering a transient state below the onset, the system may undergo phase amplification. As observed in the HCS model, for large system sizes, in the microphase separation region, one could observe more regular structures, like the nonequilibrium spatially-modulated stripes observed in the HL model. The snapshots, presented in Fig.~\ref{Fig_DomainSize_ShoulderModel} demonstrate that the off-lattice MD simulations produce nonequilibrium bicontinuous ``microemulsion'' structures. 

\section{Generalized Cahn-Hilliard Theory}
We now seek to provide a theoretical framework to model and explain the above-described phenomenology provided by simulations. To this end, we consider a generalized version \cite{MFT_PT_2021} of the classic Cahn-Hilliard theory \cite{Cahn_Hillaird_1958,cahn_phase_1965}. In this approach, we consider the source of forced interconversion to be an imbalance in chemical potentials that alters the relaxation of the interconversion dynamics to equilibrium, thereby generating a nonequilibrium steady-state condition. The imbalance of the chemical potentials can be introduced through unbalanced intermolecular forces, like in the DCM model; through an imbalance of internal energy, as in the HL model; or externally through a flux of energy-carrying agents, like in the HCS model. In the Cahn-Hilliard theory \cite{Cahn_Hillaird_1958,cahn_phase_1965}, the evolution of the local concentration of one of the alternative species, $c_\text{A}$, expressed through the order parameter $\phi = 2(c_\text{A}-1/2$), is described via the {conserved} continuity equation: $\partial\phi/\partial t = \nabla\cdot\textbf{J}_\text{C}$. In this expression, $\textbf{J}_\text{C}$ is the mutual diffusion flux, related to the gradient of the equilibrium chemical potential difference, $\hat{\mu}\equiv\mu/k_\text{B}T_\text{c}= \hat{{\mu}}_A-{\hat{\mu}}_B$, as $\textbf{J}_\text{C} = -M\nabla\hat{\mu}$, where $M$ is the molecular mobility, $k_\text{B}$ is Boltzmann's constant, and $T_\text{c}$ is the liquid-liquid critical demixing temperature. In the symmetric, regular-solution model, the chemical potential {is represented by} a sum of entropic and enthalpic mixing contributions {and a ``local'' spatially-dependent term \cite{Cahn_Hillaird_1958,cahn_phase_1965,MFT_PT_2021}}: 
\begin{equation}\label{Eq_ChemDiff_RegSol}
    \hat{\mu}= \hat{T}\ln\left(\frac{1+\phi}{1-\phi}\right)-\hat{a}\phi - R_0^2\nabla^2\phi
\end{equation}
where $\hat{a}\equiv a/k_\text{B}T_\text{c}$ is the non-ideality interaction parameter, $\hat{T}=T/T_\text{c}$ is the reduced temperature, {and $R_0$ is the microscopic length scale on the order of the molecular size}. Minimization of this equation results in the critical temperature of demixing, $T_\text{c}=a/2k_\text{B}$ \cite{Hildebrand_Regular_1962} (see Sec. 1.A in the SI). The conserved continuity equation describes the dynamics of phase separation \cite{MFT_PT_2021}. 

The generalized Cahn-Hilliard theory \cite{MFT_PT_2021} includes the flux associated with the non-conserved interconversion dynamics, $J_\text{NC}$, which contains both the natural and forced interconversions as $J_\text{NC} = -L\hat{\mu}+K\tilde{\mu}$ (see Sec. 1.B. in the SI for details). Natural interconversion is described by $-L\hat{\mu}$, while forced interconversion is described by $K\tilde{\mu}$, where $L$ and $K$ are kinetic coefficients.  {We note that the flux, $J_\text{NC}$, contains two different chemical potentials: an equilibrium chemical potential, $\hat{\mu}$, given by Eq.~(\ref{Eq_ChemDiff_RegSol}), the same for both natural interconversion and diffusion dynamics, and a nonequilibrium (``unbalanced'') chemical potential $\tilde{\mu}$.} In the first-order approximation, $\tilde{\mu}$ {is a non-local (spatially-independent) chemical potential that} scales linearly with the order parameter as $\tilde{\mu}\sim -\phi$. Thus, the general continuity equation involving all three dynamical processes is given in the form \cite{Longo_Structure_2021,MFT_PT_2021}
\begin{equation}\label{Eq_Central_Continuity}
    \frac{\partial\phi}{\partial t} = M\nabla^2\hat{\mu}-L\hat{\mu}+K\tilde{\mu}
\end{equation}
In this expression, the terms on the right hand side describe: diffusion, natural interconversion, and forced interconversion, respectively. The kinetic coefficients, $M$, $L$, and $K$, typically depend on temperature, pressure, and (for $L$ and $K$) the interconversion probability. In the lowest-order approximation, the natural interconversion dynamics scales as $-L\hat{\mu}\sim -L(\hat{T}-1)\phi$, which is positive for $T<T_\text{c}$, while in the same approximation, the forced interconversion dynamics scales as $K\tilde{\mu}\sim -K\phi$, which is negative, meaning that it always opposes natural interconversion. Consequently, the difference between the unbalanced and balanced chemical potentials, $\Delta\tilde{\mu}=\hat{\mu}-\tilde{\mu}$, provides the net driving force on the system. 

By rearranging Eq.~(\ref{Eq_Central_Continuity}) to explicitly include this driving force, the dynamics of the natural and forced interconversions may be combined into a single term with a kinetic coefficient, $L$. Redefining the unbalanced chemical potentials as $\tilde{\mu}' = (K/L)\tilde{\mu}$, a simplified continuity equation may be expressed through two dynamic processes in the form
\begin{equation}\label{Eq_Two_Dynamics}
    \frac{\partial\phi}{\partial t}=M\nabla^2\hat{\mu}-L\Delta\tilde{\mu}'
\end{equation}
where $\Delta\tilde{\mu}'$ is the difference between the balanced and (redefined) unbalanced chemical potential, $\Delta\tilde{\mu}' = \hat{\mu}-\tilde{\mu}'$, such that the second term describes the coupled natural-forced interconversion dynamics in the system. In this form Eq.~(\ref{Eq_Two_Dynamics}) is similar to the continuity equations used to describe the dynamic behavior in active matter systems \cite{Grafke_Self_Org_2017,li_non-equilibrium_2020,Kirschbaum_Biomolecular_2021,Heirarchical_Li_2021}. {However, we note that our work is different from other studies of active matter systems as we explicitly consider the evolution of the system toward equilibrium and the behavior at equilibrium. For instance, in our approach, both the natural interconversion and diffusion dynamics depend on the local (spatially-dependent) part of the chemical potential, $\hat{\mu}$ (see Sec. 1.C in the SI for details). We emphasize that this property of dynamics is inherent to all of the models simulated in this work.} 

\begin{figure*}[t]
    \centering
    \includegraphics[width=0.32\linewidth]{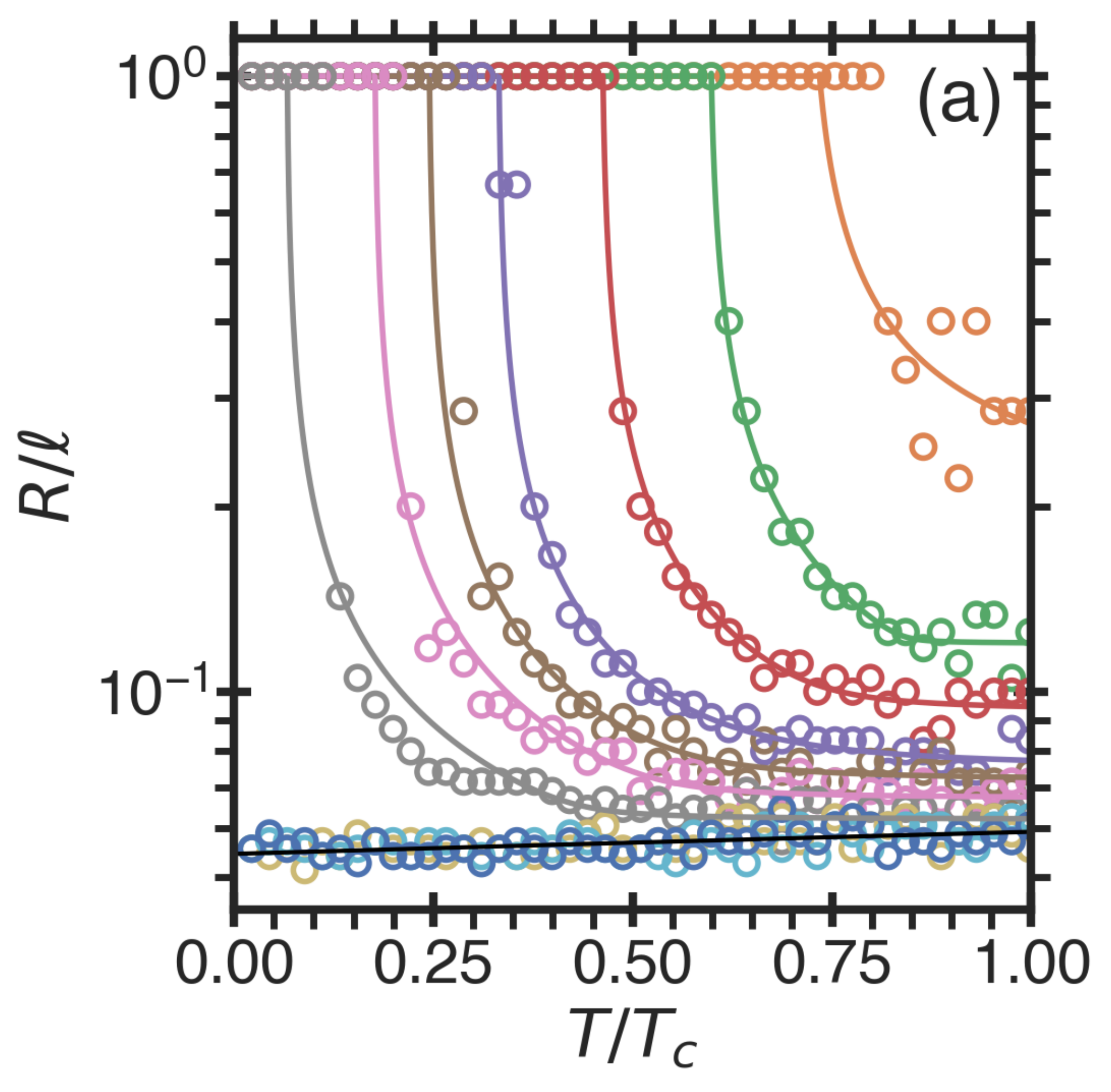}
    \includegraphics[width=0.32\linewidth]{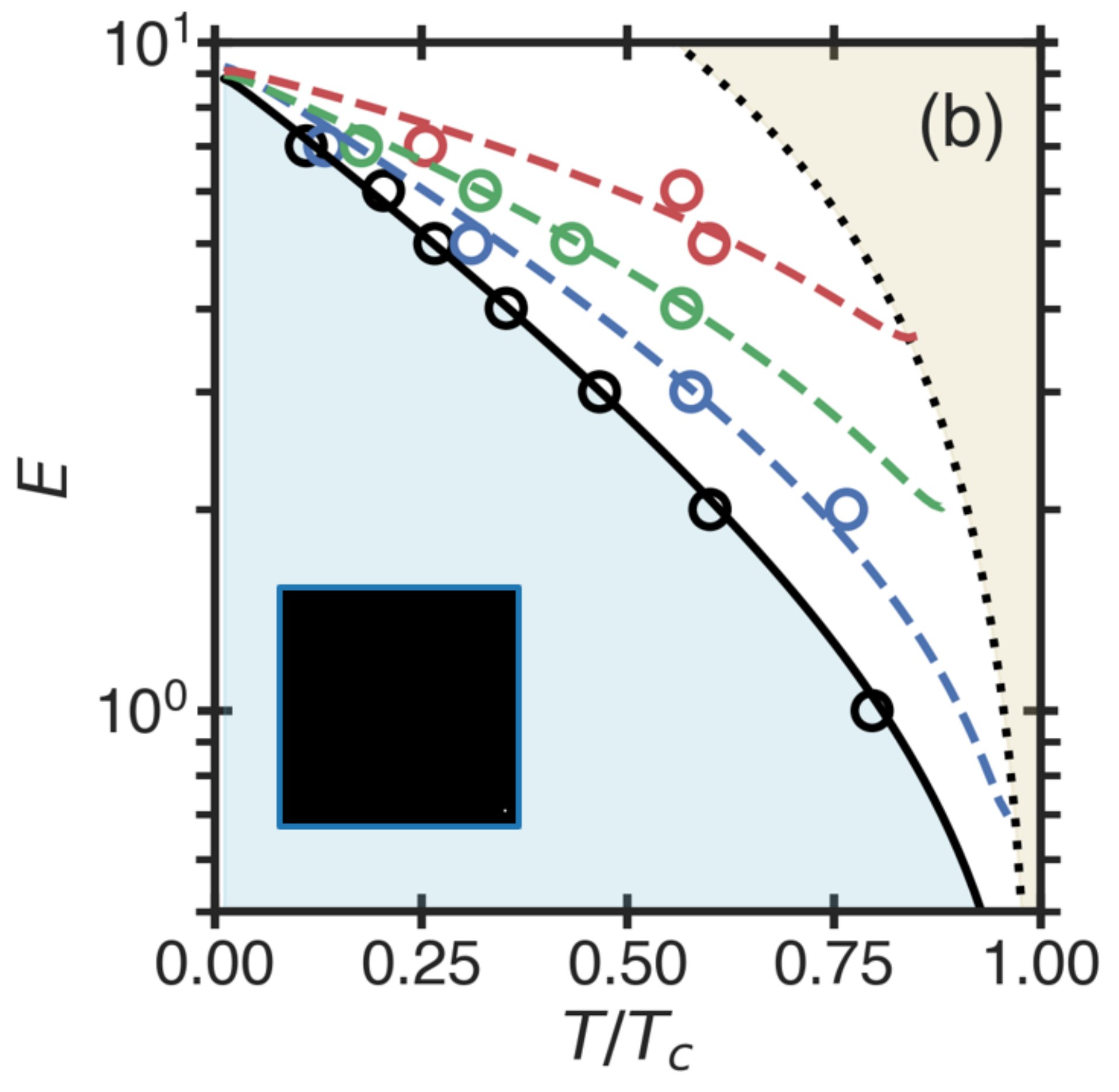}
    \includegraphics[width=0.32\linewidth]{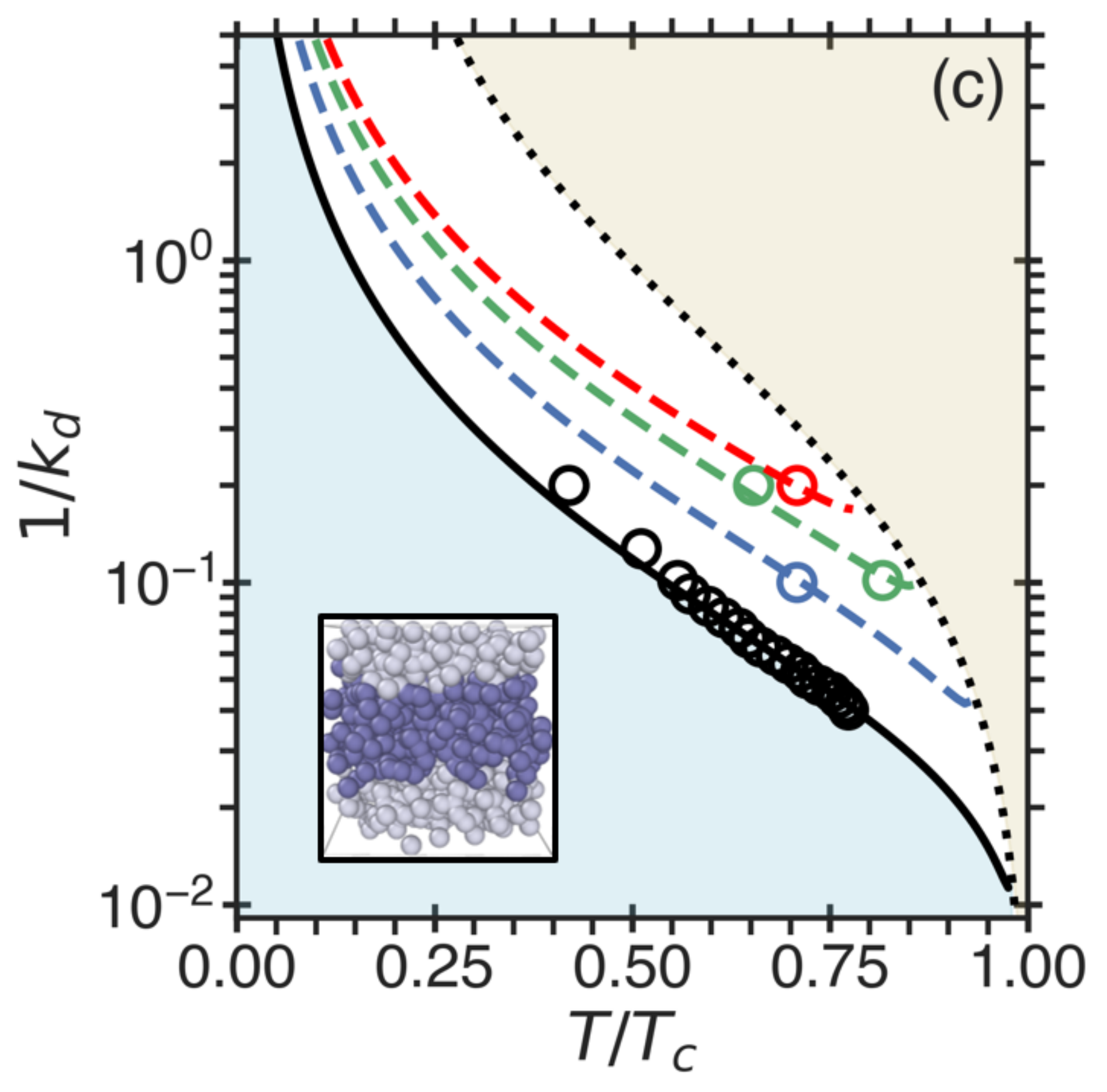}
    \caption{The onset and termination of microphase separation. (a) The steady-state domain size in the HL model for $p_r = 1/128$ and different {forced} interconversion energies from $E = 1$ (orange) to $E=10$ (dark blue) in steps of $\Delta E = 1$ (Fit parameters provided in Table~S1). For $E>E^{**}=7$ (the termination energy), the data collapse into a single line (black), indicating that the characteristic steady-state domain size remains on the order of the microscopic length scale $R_0(T)$, which corresponds to homogeneous steady-state systems for all temperatures. For $E \le 7$, the onset of microphase separation is observed at $T = T^{*}(E^*)$, where $E^*$ is the onset energy. For $T<T^*$, the steady-state domain size is equal to the size of the system, $R=\ell$. (b) The onset energy $E^*$ (black circles and curve) for the HL model as a function of temperature for $p_r = 1/128$. Colored open circles and dashed curves correspond to steady-state domain sizes: $R = 0.143$ (blue), $R = 0.095$ (green), and $R = 0.074$ (red). (c) The inverse onset rigidity parameter  $1/k_d^*$ (black circles and curve) for the DCM model at $P=0.1$. Colored circles and dashed curves correspond to steady-state domain sizes: $R = 0.32$ (blue), $R = 0.22$ (green), and $R = 0.18$ (red). In (b) and (c), the blue area corresponds to the phase separation on the scale of the simulation box, the white area corresponds to microphase separation, and the yellow area corresponds to homogeneous steady states. The images in (b) and (c) correspond to the different final states of the systems below $E^*(T^*)$ and $1/k_d^*(T^*)$ where the size of the phase domain is on the scale of the simulation box ($q^*\sim 1/\ell$). In (b), phase amplification is observed since for $p_r=1/128$ natural interconversion is relatively fast, while in (c), where natural interconversion is relatively slow for the simulated range of $k_d$, we find two-phase separation on the scale of the simulation box.}
    \label{Fig_Tstar}
\end{figure*}

In the lowest-order approximation, the unbalanced chemical potential directly impacts the enthalpy of mixing in a compressible system. We note that in an incompressible system, the heat of mixing is given by the change of the internal energy. As such, if the unbalanced chemical potential contains a tunable energy parameter, $\epsilon$, which controls the strength of the source of forced interconversion, then the difference in chemical potentials is
\begin{equation}\label{Eq_Chem_Pot_Differ}
    \Delta \tilde{\mu}' = \hat{T}\ln\left(\frac{1+\phi}{1-\phi}\right)+\left(\frac{K}{L}\epsilon-\hat{a}\right)\phi  - R_0^2\nabla^2\phi
\end{equation}
If the source of forced interconversion, $K\epsilon$, is not sufficiently strong to overcome the natural interconversion, $K\epsilon<L\hat{a}$, then equilibrium interconversion (although limited by a reduced heat of mixing) will still dominate the system and phase amplification will be observed. However, if $K\epsilon\ge L\hat{a}$, then forced interconversion will dominate the system and microphase separation will be observed. In the case when the kinetic coefficients of forced interconversion, $K$, and natural interconversion, $L$, define the same rates, $K=L$ (as is the unique case for the DCM model), then the case $\epsilon=\hat{a}$ corresponds to the specific case of the DCM model considered in \cite{Uralcan_Interconversion_2020} for the particular case $\lambda=0.5$. We note that this case is also possible in the HL and HCS models for specific combinations of $L$, $K$, and $\epsilon$. Thus, in all three models studied in this work, the source of forced interconversion can be considered as affecting the contribution from the enthalpy or internal energy of mixing ($\hat{a}\phi$), and may, either fully (as in the DCM model) or partially (as in the HL and HCS models), cancel this contribution, thereby overriding natural interconversion (see Sec. 2 in the SI for details). 

\section{Discussion}
In this section, we illustrate the ubiquitous nature of the nonequilibrium behavior in interconverting binary systems through a comparison of the simulated models. We also discuss the limiting conditions for the observation of microphase separation. We show that under certain constrains all three of the considered models, as well as the model of Glotzer \textit{et al.} \cite{glotzer_monte_1994,Glotzer_consistent_1994,glotzer_reaction-controlled_1995,Glotzer_Review_1995,Christensen_Segregation_1996,Krishnan_Molecular_2015,Lamorgese_Spinodal_2016}, would exhibit the same  behavior. 

\subsection{Comparison Between the Sources of Microphase Separation} 
As indicated by Eq.~(\ref{Eq_Central_Continuity}), there are three kinetic coefficients, $L$, $M$, and $K$, whose interplay determines whether microphase separation may occur. In the HL model, the kinetic coefficients, $L$ and $M$ (considered to be independent of temperature and pressure), determine the probability of natural interconversion, $p_r$, through $p_r = L/(Mq^2 + L)$ \cite{Shum_Phase_2021,MFT_PT_2021}. Thus, $M=0$ corresponds to fast interconversion ($p_r=1$), while $L = 0$ corresponds to no natural interconversion ($p_r =0$). In this model, for small $p_r$, we approximate the source of forced interconversion as being uncoupled from natural interconversion and related to the kinetic coefficient $K$ as $K\propto E^2$, where the prefactor depends on temperature only \cite{Longo_Structure_2021}. In the limit when $E\to E_\text{max}=12$, the HL model becomes equivalent to the model of Glotzer \textit{et al.}\cite{glotzer_monte_1994,Glotzer_consistent_1994,glotzer_reaction-controlled_1995,Glotzer_Review_1995,Christensen_Segregation_1996,Krishnan_Molecular_2015,Lamorgese_Spinodal_2016}, in which there is no natural interconversion ($p_r=0$).

In the DCM model, the source of forced interconversion is coupled to the natural interconversion rate through the dihedral force constant, $k_d$. The behavior of the system with different dihedral-force constants may be related to the behavior of the HL model system for different interconversion probabilities, $p_r$, by considering an interpolation between two limits as $k_d = \sqrt{(1/p_r)-1}$, such that $k_d\to 0$ ($p_r =1$) corresponds to fast natural and {forced} interconversion, while $k_d\to\infty$ ($p_r =0$) corresponds to no interconversion \cite{MFT_PT_2021}. This specific feature of the DCM model, that the natural and forced interconversions are controlled by a single parameter $k_d$, means that without interconversion (only in the limit $k_d\to \infty$), the DCM system is in equilibrium and exhibits regular phase separation. This effect is utilized in the theoretical model, Eq.~(\ref{Eq_Central_Continuity}), when both the natural and forced interconversions are controlled by the same kinetic coefficient, such that $L = K$, where the dissipative intramolecular forces may be expressed, in the interconversion dynamics, through the difference between the balanced and imbalanced chemical potentials, $\Delta\hat{\mu}$ (given by Eq.~(\ref{Eq_Chem_Pot_Differ}) where $\epsilon = \hat{a}$). The mobility, $M$ is described by the Stokes-Einstein relation, while the interconversion kinetic coefficient, $L$, has been found to be proportional to $M$, such that $L=MT^2/k_d^2$ \cite{Uralcan_Interconversion_2020}. 

Just like the behavior of the HL model, the source of forced interconversion in the HCS model depends on the relation between $L$ and $K$, and, in the first-order approximation, {they} may be assumed to be uncoupled from each other. Consequently, the behavior of the nonequilibrium HCS model may be described via a similar dynamic equation as used to describe the behavior of the HL model. However, in the nonequilibrium HCS model, while the natural interconversion rate is determined by the interaction cross-section of the molecules and energy-carrying agents, $L\propto b^2$, as described in Sec. 2.C in the SI, the effect of forced interconversion varies with the strength of the energy source, $\epsilon$. This effect is introduced into the difference between the balanced and imbalanced chemical potentials, Eq.~(\ref{Eq_Chem_Pot_Differ}), as a tunable parameter, such that when $\epsilon\to 0$ the nonequilibrium chemical potential $\tilde{\mu}\to 0$ and the system evolves according to equilibrium conditions. However, in the limiting case, when $K$ and $L$ are of equal magnitude and $\epsilon = \hat{a}$, such that the enthalpic contribution to the chemical potential is completely compensated, then the HCS model will be dynamically equivalent to the DCM model. 

For all of the models considered, in the first-order approximation, the domain size ($R$) scales with interconversion probability as $R\propto 1/\sqrt{p_r}\propto k_d\propto 1/b$. The steady-state domain size also depends on the temperature. In the DCM and HCS models, this temperature dependence originates from $M$ and $L$, while in the HL model, this temperature dependence originates from the relationship between $K$ and $E$. As shown by the solid curves on Figs.~\ref{Fig_Theory_NoneqHybrid}-\ref{Fig_DomainSize_ShoulderModel}, we obtain a quantitative comparison between the simulation results and the theory. For more details, see Sec. 2 in the SI.

\subsection{Onset and Termination of Microphase Separation}
As observed in the simulations of all three microscopic models, there are three regions in which different phenomena may be observed. They may be identified by the interplay between diffusion, natural interconversion, and forced interconnection, which are described by the kinetic coefficients $M$, $L$, and $K$ in Eq.~(\ref{Eq_Central_Continuity}). When natural interconversion, $L$, is faster that the diffusion or forced interconversion rate, then phase amplification is observed. For instance, in the HL model this occurs where $T<T^*$ and $E<E^*$. When the diffusion rate, $D\propto Mq^2$, is faster than the natural and the forced interconversion rates, then transient (``apparent'') two-phase separation on the scale of the simulation box is observed. For instance, in the DCM and HCS models, this is observed where $T<T^*$ and $1/k_d < 1/k_d^*$ (DCM) or $b<b^*$ (HCS). The curve that separates the phase amplification or transient two-phase region from the microphase separation region is referred to as the onset. This curve may be found from Eq.~(\ref{Eq_Central_Continuity}), considered for the particular case, when the characteristic size of the phase domains, determined from the maximum of the growth rate equation (see Eq.~S6 for details), becomes equal to the size of the simulation box, $q\sim 1/\ell$. 

Alternatively, when forced interconversion, $K$, is faster than diffusion and natural interconversion, then the alternative species will interconvert so fast that no dissipative structures may form and only a homogeneous steady state with statistically equal concentration of the interconverting species will remain. In this case, the characteristic size of the domains is of the order of the molecular length scale, $R_0$. We define the temperature and energy at which this occurs as the ``termination'' of microphase separation ($T^{**}$ and $E^{**}$). For instance in the HL {and HCS} models, this occurs when $E>E^{**}$ {($\epsilon > \epsilon^{**}$) or} $T>T^{**}$, while in the DCM model, since the natural and forced interconversion are coupled, this region occurs when $1/k_d>1/k_d^{**}$ {or} $T>T^{**}$. This effect is shown in Fig.~\ref{Fig_Tstar}a where the characteristic size for $E>7$ is $R_0$ for all temperatures. An increase in $R_0$ with temperature could be attributed to the growing correlation length of concentration fluctuations upon the approach to the critical temperature. The curve that separates the microphase region from the homogeneous steady-state region is referred to as the termination curve, and may be found in the present theory, when the maximum of the characteristic growth rate with respect to $q$ reaches zero (see Eq.~S6 for details).

The region of the phase diagram where microphase separation occurs (between the onset and the termination {lines}) is where diffusion, natural interconversion, and forced interconversion are balanced and where steady-state dissipative structures are observed. The characteristic length scale of the microphase region is predominantly given by the interplay between diffusion and forced interconversion (see Sec. 1.B in the SI for details). The comparison between these three regions in the HL and DCM models is illustrated in Figs.~\ref{Fig_Tstar}(b and c). As shown, the onset and termination curves behave similarly between these regions.

We have shown that under certain limits all of the simulated models would exhibit identical dynamic behavior. These limits are summarized as: 1) The limit of complete phase separation occurs when $p_r\to 0$ and $E\to 0$ (HL), $k_d\to\infty$ (DCM), and $b\to 0$ and $\varepsilon\to 0$ (HCS). 2) Microphase separation in the absence of natural interconversion occurs in the HL {and HCS models} when $E \le E_\text{max}$ {and $\varepsilon \le \varepsilon_\text{max}$} {at constant $p_r$ or $b$ and $T$}. {In the} limit {when $p_r$ and $b$ are small, while $E \ge E_\text{max}$ and $\varepsilon \ge \varepsilon_\text{max}$,} the dynamic behavior of the HL {and HCS models} becomes {equivalent} to the model of Glotzer \textit{et al.}  \cite{glotzer_monte_1994,Glotzer_consistent_1994,glotzer_reaction-controlled_1995,Glotzer_Review_1995,Christensen_Segregation_1996,Krishnan_Molecular_2015,Lamorgese_Spinodal_2016}. 3) The dynamic behavior of the DCM model (imbalance of interaction forces) is a limiting case of the behavior of the {HL and} HCS {models} (external source of energy-carrying particles). 4) Limit of a homogeneous steady state occurs in all the models for $T>T^{**}$, when {$E > E^{**}$} (HL), $k_d\to 0$ (DCM), and {$\varepsilon > \varepsilon^{**}$} (HCS). 

\section{Summary and Applications}
In summary, depending on the rate of interconversion and distance to the critical temperature, there are three possible scenarios that are observed in the behavior of our three microscopic models of mixtures that exhibit phase separation in the presence of both the natural and forced interconversions of species: (1) phase amplification or transient two-phase separation on the scale of the simulation box, (2) microphase separation, and (3) homogeneous steady state. Unlike the modulated phases and bicontinuous microemulsion structures observed in equilibrium conditions or the patterns formed in ``frozen'' metastable conditions \cite{Gompper_Lattice_1994,Andelman_Modulated_2009,Jones_Soft_Matter_2002,Borukhov_Polyelectrolyte_2000,anisimov_thermodynamics_2018,li_non-equilibrium_2020}, the steady-state dissipative structures investigated in this work are the result of the continuous energy supply to the system. The three physically different models considered in this work, as well as the model of Glotzer \textit{et al.}  \cite{glotzer_monte_1994,Glotzer_consistent_1994,glotzer_reaction-controlled_1995,Glotzer_Review_1995,Christensen_Segregation_1996,Krishnan_Molecular_2015,Lamorgese_Spinodal_2016}, demonstrate identical behavior under certain limits. This behavior can be quantitatively unified by the phenomenological model of phase transitions affected by molecular interconversion \cite{MFT_PT_2021}. 

Nonequilibrium microphase separation could exist in a wide class of systems, including ``active matter systems,'' a recent focus of theoretical and experimental studies \cite{Weber_Active_2019,Ramaswamy_Active_2010}, as well as biomolecular condensates (\textit{e.g.} membraneless organelles), where natural  interconversion could be caused by mechanisms like polymerization, protein folding-unfolding, and self-assembly, while {forced} interconversion could be generated by the nonequilibrium environment of the cell \cite{Ramaswamy_Active_2010,kim_mutations_2013,Lin_Condensates_2018,Shakhnovich_Organelles_2019,Pappu_Condensates_2020,Kirschbaum_Biomolecular_2021,Heirarchical_Li_2021,Pappu_Ligand_2021,Pappu_Conceptual_2022,Shakhnovich_MultiDomain_2022,Feric_mtCondensates_2022}. The developed approach could be applicable to understanding and quantitatively describing these phenomena. In addition, microphase separation may also exist in other supramolecular structures, e.g. polymer solutions in the presence of photochemical reactions \cite{Kim_Metallic_2013,wang_progress_2019}. Our approach could be linked to other dissipative phenomena, like hydrodynamic instabilities and bifurcations in chemical reactions \cite{nicolis_self-organization_1977}.


\acknow{The authors thank Mihail Popescu for useful discussions. This work is a part of the research collaboration between the University of Maryland, Princeton University, Boston University, and Arizona State University supported by the National Science Foundation. S.V.B., M.A.A., and P.G.D. acknowledge the financial support of the National Science Foundation (Award Nos. CHE-1856496, CHE-1856479, and CHE-1856704, respectively). S.V.B. acknowledges the partial support of this research through Bernard W. Gamson Computational Science Center at Yeshiva College. {B.U. acknowledges the partial computational support of Bogazici University Research Fund 17841 and TUBITAK T118C220.} Some simulations were performed on computational resources managed and supported by Princeton Research Computing, a consortium of groups including the Princeton Institute for Computational Science and Engineering (PICSciE) and the Office of Information Technology’s High Performance Computing Center and Visualization Laboratory at Princeton University.}

\showacknow{} 

\bibliography{reference}

\begin{thebibliography}{10}

\bibitem{Vvedensky_Transformations_2019}
DD Vvedensky, {\em Transformations of Materials}.
\newblock (Morgan and Claypool Publishers, San Rafael, California), (2019).

\bibitem{Fultz_Phase_2020}
B Fultz, {\em Phase Transitions in Materials}.
\newblock (Cambridge University Press, Cambridge, UK), 2 edition, (2020).

\bibitem{Kim_Metallic_2013}
DH Kim, WT Kim, ES Park, N Mattern, J Eckert, Phase separation in metallic
  glasses.
\newblock {\em\protect\JournalTitle{Prog. Matter. Sci.}} \textbf{58},
  1103--1172 (2013).

\bibitem{Lin_Nanocrystallization_2015}
C Lin, C Bocker, C Russel, Nanoscrystallization in oxyfluoride glasses
  controlled by amorphous phase separation.
\newblock {\em\protect\JournalTitle{Nano Lett.}} \textbf{15}, 6764--6769
  (2015).

\bibitem{Gompper_Lattice_1994}
G Gompper, M Schick, Lattice theories of microemulsions in {\em Micelles,
  Membranes, Microemulsions, and Monolayers}, eds.{} WM Gelbart, A Ben-Shaul, D
  Roux.
\newblock (Springer, New York, NY), 1 edition, pp. 395--426 (1994).

\bibitem{Andelman_Modulated_2009}
D Andelman, RE Rosensweig, Modulated phases: Review and recent results.
\newblock {\em\protect\JournalTitle{J. Phys. Chem. B}} \textbf{113}, 3785--3798
  (2009).

\bibitem{Jones_Soft_Matter_2002}
RAL Jones, {\em Soft Condensed Matter}.
\newblock (Oxford University Press, Oxford, UK), (2002).

\bibitem{Borukhov_Polyelectrolyte_2000}
I Borukhov, D Andelman, MC Regis~Borrega, L Leibler, H Orland, Polyelectrolyte
  titration: Theory and experiment.
\newblock {\em\protect\JournalTitle{J. Phys. Chem. B}} \textbf{104},
  11027--11034 (2000).

\bibitem{anisimov_thermodynamics_2018}
MA Anisimov, et~al., Thermodynamics of {Fluid} {Polyamorphism}.
\newblock {\em\protect\JournalTitle{Phys. Rev. X}} \textbf{8}, 011004 (2018).

\bibitem{li_non-equilibrium_2020}
YI Li, ME Cates, Non-equilibrium phase separation with reactions: a canonical
  model and its behaviour.
\newblock {\em\protect\JournalTitle{J. Stat. Mech. Theory Exp.}} \textbf{2020},
  053206 (2020).

\bibitem{Ricci_Computational_2013}
F Ricci, FH Stillinger, PG Debenedetti, A computational investigation of
  attrition-enhanced chiral symmetry breaking in conglomerate crystals.
\newblock {\em\protect\JournalTitle{J. Chem. Phys.}} \textbf{139} (2013).

\bibitem{Shum_Phase_2021}
NA Shumovskyi, TJ Longo, SV Buldyrev, MA Anisimov, Phase amplification in
  spinodal decomposition of immiscible fluids with interconversion of species.
\newblock {\em\protect\JournalTitle{Phys. Rev. E}} \textbf{103}, L060101
  (2021).

\bibitem{MFT_PT_2021}
TJ Longo, MA Anisimov, Phase transitions affected by natural and forceful
  molecular interconversion.
\newblock {\em\protect\JournalTitle{J. Chem. Phys.}} \textbf{156}, 084502
  (2022).

\bibitem{Miyata_PolyemrMix_2017}
Q Tran-Cong-Miyata, H Nakanishi, Phase separation of polymer mixtures driven by
  photochemical reactions: current status and perspectives.
\newblock {\em\protect\JournalTitle{Polym. Int.}} \textbf{66}, 213--222 (2017).

\bibitem{Weber_Active_2019}
CA Weber, D Zwicker, F J\"ulicher, CF Lee, Physics of active emulsions.
\newblock {\em\protect\JournalTitle{Rep. Prog. Phys.}} \textbf{82}, 064601
  (2019).

\bibitem{Boeynaems_Protein_2018}
S Boeynaems, et~al., Protein phase separation: A new phase in cell biology.
\newblock {\em\protect\JournalTitle{Trends Cell Biol.}} \textbf{28}, 420--435
  (2018).

\bibitem{Hyman_LLBiology_2014}
AA Hyman, CA Weber, F Jülicher, Liquid-liquid phase separation in biology.
\newblock {\em\protect\JournalTitle{Annu. Rev. Cell Dev.}} \textbf{30}, 39--58
  (2014).

\bibitem{huberman_striations_1976}
BA Huberman, Striations in chemical reactions.
\newblock {\em\protect\JournalTitle{J. Chem. Phys.}} \textbf{65}, 2013--2019
  (1976).

\bibitem{Verdasca_Chemically_1995}
J Verdasca, P Borckmans, G Dewel, Chemically frozen phase separation in an
  adsorbed layer.
\newblock {\em\protect\JournalTitle{Phys. Rev. E}} \textbf{52}, R4616--R4619
  (1995).

\bibitem{carati_chemical_1997}
D Carati, R Lefever, Chemical freezing of phase separation in immiscible binary
  mixtures.
\newblock {\em\protect\JournalTitle{Phys. Rev. E}} \textbf{56}, 3127--3136
  (1997).

\bibitem{glotzer_monte_1994}
SC Glotzer, D Stauffer, N Jan, Monte {Carlo} simulations of phase separation in
  chemically reactive binary mixtures.
\newblock {\em\protect\JournalTitle{Phys. Rev. Lett.}} \textbf{72}, 4109--4112
  (1994).

\bibitem{Glotzer_consistent_1994}
SC Glotzer, A Coniglio, Self-consistent solution of phase separation with
  competing interactions.
\newblock {\em\protect\JournalTitle{Phys. Rev. E}} \textbf{50}, 4241--4244
  (1994).

\bibitem{glotzer_reaction-controlled_1995}
SC Glotzer, EA Di~Marzio, M Muthukumar, Reaction-{Controlled} {Morphology} of
  {Phase}-{Separating} {Mixtures}.
\newblock {\em\protect\JournalTitle{Phys. Rev. Lett.}} \textbf{74}, 2034--2037
  (1995).

\bibitem{lefever_comment_1995}
R Lefever, D Carati, N Hassani, Comment on ``{Monte} {Carlo} {Simulations} of
  {Phase} {Separation} in {Chemically} {Reactive} {Binary} {Mixtures}''.
\newblock {\em\protect\JournalTitle{Phys. Rev. Lett.}} \textbf{75}, 1674--1674
  (1995).

\bibitem{Glotzer_Response_1995}
SC Glotzer, D Stauffer, N Jan, Glotzer, stauffer, and jan reply:.
\newblock {\em\protect\JournalTitle{Phys. Rev. Lett.}} \textbf{75}, 1675--1675
  (1995).

\bibitem{Glotzer_Review_1995}
SC Glotzer, {\em Computer Simulations of Spinodal Decomposition in Polymer
  Blends}.
\newblock pp. 1--46 (1995).

\bibitem{Christensen_Segregation_1996}
JJ Christensen, K Elder, HC Fogedby, Phase segregation dynamics of a chemically
  reactive binary mixture.
\newblock {\em\protect\JournalTitle{Phys. Rev. E}} \textbf{54}, R2212--R2215
  (1996).

\bibitem{Krishnan_Molecular_2015}
R Krishnan, S Puri, Molecular dynamics study of phase separation in fluids with
  chemical reactions.
\newblock {\em\protect\JournalTitle{Phys. Rev. E}} \textbf{92}, 052316 (2015).

\bibitem{Lamorgese_Spinodal_2016}
A Lamorgese, R Mauri, Spinodal decomposition of chemically reactive binary
  mixtures.
\newblock {\em\protect\JournalTitle{Phys. Rev. E}} \textbf{94}, 022605 (2016).

\bibitem{Uralcan_Interconversion_2020}
B Uralcan, TJ Longo, MA Anisimov, FH Stillinger, PG Debenedetti,
  Interconversion-controlled liquid-liquid phase separation in a molecular
  chiral model.
\newblock {\em\protect\JournalTitle{J. Chem. Phys.}} \textbf{155}, 204502
  (2021).

\bibitem{Longo_Structure_2021}
TJ Longo, NA Shumovskyi, SM Asadov, SV Buldyrev, MA Anisimov, Structure factor
  of a phase separating binary mixture with natural and forceful
  interconversion of species.
\newblock {\em\protect\JournalTitle{J. Non-Cryst. Solids: X}} \textbf{13},
  100082 (2022).

\bibitem{Heirarchical_Li_2021}
YI Li, ME Cates, Hierarchical microphase separation in non-conserved active
  mixtures.
\newblock {\em\protect\JournalTitle{Eur. Phys. J. E}} \textbf{44}, 119 (2021).

\bibitem{Latinwo_MolecModel_2016}
F Latinwo, FH Stillinger, PG Debenedetti, Molecular model for chirality
  phenomena.
\newblock {\em\protect\JournalTitle{J. Chem. Phys.}} \textbf{145}, 154503
  (2016).

\bibitem{Metropolis_Basic_1963}
N Metropolis, RL Ashenhurst, Basic {Operations} in an {Unnormalized}
  {Arithmetic} {System}.
\newblock {\em\protect\JournalTitle{IEEE Trans. Comput.}} \textbf{EC-12},
  896--904 (1963).

\bibitem{Petsev_Effect_2021}
ND Petsev, FH Stillinger, PG Debenedetti1, Effect of configuration-dependent
  multi-body forces on interconversion kinetics of a chiral tetramer model.
\newblock {\em\protect\JournalTitle{J. Chem. Phys.}} \textbf{155}, 084105
  (2021).

\bibitem{Heuer_Critical_1993}
HO Heuer, Critical crossover phenomena in disordered ising systems.
\newblock {\em\protect\JournalTitle{J. Phys. A: Math. Gen.}} \textbf{26},
  L333--L339 (1993).

\bibitem{Alder_Studies_1959}
BJ Alder, TE Wainwright, Studies in molecular dynamics. {I}. general method.
\newblock {\em\protect\JournalTitle{J. Chem. Phys.}} \textbf{31}, 459 (1959).

\bibitem{Rapaport_Art_2004}
DC Rapaport, {\em The Art of Molecular Dynamics Simulation}.
\newblock (Cambridge University Press, Cambridge, UK), 2 edition, (2004).

\bibitem{Berendsen_Molecular_1984}
HJC Berendsen, JPM Postma, WF van Gunsteren, A DiNola, JR Haak, Molecular
  dynamics with coupling to an external bath.
\newblock {\em\protect\JournalTitle{J. Chem. Phys.}} \textbf{81}, 3684 (1984).

\bibitem{Cahn_Hillaird_1958}
JW Cahn, JE Hilliard, Free energy of a nonuniform system. {I}. interfacial free
  energy.
\newblock {\em\protect\JournalTitle{J. Chem. Phys.}} \textbf{28}, 258 (1958).

\bibitem{cahn_phase_1965}
JW Cahn, Phase {Separation} by {Spinodal} {Decomposition} in {Isotropic}
  {Systems}.
\newblock {\em\protect\JournalTitle{J. Chem. Phys.}} \textbf{42}, 93--99
  (1965).

\bibitem{Hildebrand_Regular_1962}
J Hildebrand, R Scott, {\em Regular Solutions}, Prentice-Hall international
  series in chemistry.
\newblock (Prentice-Hall), (1962).

\bibitem{Grafke_Self_Org_2017}
T Grafke, ME Cates, E Vanden-Eijnden, Spatiotemporal self-organization of
  fluctuating bacterial colonies.
\newblock {\em\protect\JournalTitle{Phys. Rev. Lett.}} \textbf{119}, 188003
  (2017).

\bibitem{Kirschbaum_Biomolecular_2021}
J Kirschbaum, D Zwicker, Controlling biomolecular condensates via chemical
  reactions.
\newblock {\em\protect\JournalTitle{J. R. Soc. Interface}} \textbf{18},
  20210255 (2021).

\bibitem{Ramaswamy_Active_2010}
S Ramaswamy, The mechanics and statistics of active matter.
\newblock {\em\protect\JournalTitle{Annu. Rev. Condens. Matter Phys.}}
  \textbf{1}, 323--345 (2010).

\bibitem{kim_mutations_2013}
HJ Kim, et~al., Mutations in prion-like domains in {hnRNPA2B1} and {hnRNPA1}
  cause multisystem proteinopathy and {ALS}.
\newblock {\em\protect\JournalTitle{Nature}} \textbf{495}, 467--473 (2013).

\bibitem{Lin_Condensates_2018}
YH Lin, JD Forman-Kay, HS Chan, Theories for sequence-dependent phase behaviors
  of biomolecular condensates.
\newblock {\em\protect\JournalTitle{Biochemistry}} \textbf{57}, 2499--2508
  (2018).

\bibitem{Shakhnovich_Organelles_2019}
R Ranganathan, E Shakhnovich, Dynamic metastable long-living droplets formed by
  sticker-spacer proteins.
\newblock {\em\protect\JournalTitle{eLife}} \textbf{9}, e56159 (2020).

\bibitem{Pappu_Condensates_2020}
F Dar, R Pappu, Phase separation: Restricting the sizes of condensates.
\newblock {\em\protect\JournalTitle{eLife}} \textbf{9}, e59663 (2020).

\bibitem{Pappu_Ligand_2021}
KM Ruff, F Dar, RV Pappu, Ligand effects on phase separation of multivalent
  macromolecules.
\newblock {\em\protect\JournalTitle{Proc. Natl. Acad. Sci.}} \textbf{118},
  e2017184118 (2021).

\bibitem{Pappu_Conceptual_2022}
T Mittag, RV Pappu, A conceptual framework for understanding phase separation
  and addressing open questions and challenges.
\newblock {\em\protect\JournalTitle{Mol. Cell}} \textbf{82}, 2201--2214 (2022).

\bibitem{Shakhnovich_MultiDomain_2022}
S Ranganathan, E Shakhnovich, The physics of liquid-to-solid transitions in
  multi-domain protein condensates.
\newblock {\em\protect\JournalTitle{Biophysical Journal}} \textbf{121},
  2751--2766 (2022).

\bibitem{Feric_mtCondensates_2022}
M Feric, et~al., Mesoscale structure-function relationships in mitochondrial
  transcriptional condensates.
\newblock {\em\protect\JournalTitle{Proc. Natl. Acad. Sci.}} \textbf{119},
  e2207303119 (2022).

\bibitem{wang_progress_2019}
F Wang, et~al., Progress {Report} on {Phase} {Separation} in {Polymer}
  {Solutions}.
\newblock {\em\protect\JournalTitle{Adv. Mater.}} \textbf{31}, 1806733 (2019).

\bibitem{nicolis_self-organization_1977}
G Nicolis, I Prigogine, {\em Self-{Organization} in {Nonequilibrium} {Systems}:
  {From} {Dissipative} {Structures} to {Order} through {Fluctuations}}.
\newblock (Wiley, New York), 1 edition, (1977).

\end{thebibliography}

\end{document}